\documentclass[12pt,a4paper]{article}
\begin{document}
\begin{center} 
{\bf On the supersymmetric nonlinear evolution equations }
\vskip 0.5 cm
{\small Amitava Choudhuri$^a$, B. Talukdar$^a$ and S. Ghosh$^b$}\\
{\small $^a$\it Department of Physics, Visva-Bharati University, Santiniketan 731235, India}
{\small $^b$\it Patha Bhavana, Visva-Bharati University, Santiniketan 731235, India }\\
\vskip 0.2 cm
{\small \it e-mail : binoy123@bsnl.in}
\end{center}
\vskip 1.0 cm
{\bf Abstract.} Supersymmetrization of a nonlinear evolution equation in which the bosonic equation is independent of the fermionic variable and the system is linear in fermionic field goes by the name B-supersymmetrization. This special type of supersymmetrization plays a role in superstring theory. We provide B-supersymmetric extension of a number of quasilinear and fully nonlinear evolution equations and find that the supersymmetric system follows from the usual action principle while the bosonic and fermionic equations are individually non Lagrangian in the field variable. We point out that B-supersymmetrization can also be realized using a generalized Noetherian symmetry such that the resulting set of Lagrangian symmetries coincides with symmetries of the bosonic field equations. This observation provides a basis to associate the bosonic and fermionic fields with the terms of bright and dark solitons. The interpretation sought by us has its origin in the classic work of Bateman who introduced a reverse-time system with negative friction to bring the linear dissipative systems within the framework of variational principle.
\vskip 1.0 cm
\noindent{\bf\large 1. Introduction}
\vskip 0.3 cm
In the theories of elementary particles supersymmetry aims at a unified description of fermions and bosons i.e. of matter and interaction {[}1{]}. Understandably, construction of supersymmetric theories in this case will involve consideration of particles of different spin and of different statistics. As opposed to this, supersymmetrization of integrable nonlinear evolution equations proceeds by considering only the space supersymmetric invariance {[}2{]}. More specifically, in a superspace approach to $(1+1)$ dimensional systems the commuting space variable $x$ is extended to a doublet $(x,\,\theta)$, where $\theta$ is an anticommuting variable of Grassmann type such that $\theta^2=0$. A superfield $F$ regarded as a function of $x$ and $\theta$ has a very simple Taylor expansion in terms of $\theta$ such that
$$F(x,\,\theta)=v(x)+ \theta u(x)\,\,.\eqno(1)$$
Here $u(x)$ and $v(x)$ are the component fields of $F(x,\,\theta)$ with $v(x)$, the superpartner of $u(x)$ and the conversly. The function $F(x,\,\theta)$ is fermionic (bosonic) if $u(x)$ is bosonic (fermionic).
\par A supersymmetric system for a given evolution equation is constructed in a way that will make it invariant under the transformations $x\rightarrow x-\eta\theta$ and $\theta\rightarrow \theta+\eta$ with $\eta$, an anticommuting parameter. Since the supersymmetric transformation does not depend on time we have suppressed $t$ dependence in writing $(1)$.\par For any function $\Phi$ of the doublet  $(x,\,\theta)$ it is easy to see that
$$\delta \Phi=\eta(\partial_\theta-\theta\partial_x)\Phi\,\,.\eqno(2)$$
Thus 
$$Q=\partial_\theta-\theta\partial_x\eqno(3)$$
is the generator of the supersymmetric transformation. Using $\Phi=F(x,\,\theta)$ from $(1)$ in $(2)$ we get the transformations for the component fields as
$$\delta v=\eta u\,\,\,\,\,\,{\rm and}\,\,\,\,\,\,\delta u=\eta v_x\,\,.\eqno(4)$$
The results in $(4)$ give the so-called supersymmetry transformation since the first equation relates the bosonic field to a fermionic field while the second one does the opposite. In other words, the supersymmetry transformations regarded as transformations in superspace exhibit the fermi-bose symmetry. Two successive supersymmetry transformations lead to 
$$\delta^2 v=\eta^2v_x\,\,\,\,\,\,{\rm and}\,\,\,\,\,\,\delta^2 u=\eta^2u_x\,\,.\eqno(5)$$
This shows that a supersymmetry transformation is a sort of square root of an ordinary translation. The derivative written as
$$D=\theta \partial_x+\partial_\theta\eqno(6)$$
anticommute with $Q$ and is called the covariant superderivative presumably because expressions written in terms of $D$ and superfield are manifestly invariant under the supersymmetry transformation $(4)$. It is of interest to note that
$$D^2=\partial_x\,\,.\eqno(7)$$
\par The Korteweg-de Vries (KdV) equation {[}3{]} 
$$u_t=-u_{3x}+6uu_x\,\,.\eqno(8)$$
represents the first nonlinear evolution equation that could account for soliton formation {[}4{]}. The KdV equation can be extended to a supersymmetric system by rewriting the KdV equation in terms of the superfield and covariant derivative. This is achieved by multiplying $(8)$ with $\theta$ and then supersymmetrizing the result. Thus we have
$$F_t=-D^6F+6DFD^2F\eqno(9)$$
From $(1)$ and $(9)$ we obtain equations for the bosonic and fermionic fields as
$$u_t=-u_{3x}+6uu_x\eqno(8)$$
and
$$v_t=-v_{3x}+6uv_x\,\,.\eqno(8^\prime)$$
We note that $\theta(6uu_x)$ can be regarded as a fermionic part of either $3D^2(FDF)$ or $6DFD^2F$. Therefore, in order to construct the most general possible supersymmetric extension of the KdV equation, one must consider a linear combination of these two terms. In writing $\theta(6uu_x)$ we have not taken this point into consideration but chosen to work with only the term $6DFD^2F$. This resulted in a set of simple equations given in $(8)$ and $(8^\prime)$. Clearly, the bosonic equation does not depend on the fermionic variable and moreover, the system is linear in the fermionic field. This kind of supersymmetric extension was originally discarded as being a `trivial' supersymmetrization {[}2{]}. However, it generated a lot of interest after it was realized that equations like $(8^\prime)$ arise in the study of superstring theory {[}5{]}. A supersymmetric extension in which bosonic equation do not change in the presence of fermions has now been given the name B-supersymmetrization.
\par The aim of the present work is to provide a Lagrangian realization for the pair of equations which arise from B-supersymmetrization of nonlinear evolution equations. Our equations of interest are the modified KdV (mKdV) {[}6{]}, Hunter-Saxton (HS) {[}7{]} and Camassa-Holm (CH) {[}8{]} equations. These equations are related to the celebrated KdV equation in someway or others. In this context we note that the KdV and mKdV equation are quasilinear in the sense that their dispersive terms are linear while the HS and CH equations are fully nonlinear because their dispersive terms are nonlinear. Besides Hunter-Saxton and Camassa-Holm  equations there exist another class of fully nonlinear evolution equations (FNE) introduced by Rosenau and Hyman (RH) {[}9{]}. These equation is apparently not related to the KdV equation. In section $2$ we outline how the mKdV, Hunter-Saxton and Camassa-Holm equations are related to the KdV equation and study their supersymmetric structure. We also try to present a similar supersymmetric generalization of the third-order RH equation. In section $3$ we contemplate deriving all these supersymmetric equations from the action principle.\par A single evolution equation is never an Euler-Lagrange expression. We need to couple a given equation with an associated one to construct a physically complete system in the sense of the variational principle. As an interesting curiosity we note that the associated equation is the fermionic partner of the original bosonic equation. The techanique of introducing an auxiliary equation for variational formulation of physical problems has an old root in the classical mechanics literature. For example, in a celebrated work Bateman {[}10{]} noted that a dissipative system is physically incomplete and one needs an additional equation when one attempts to derive the defining equations from an action principle.  In section $4$ we make some concluding remarks. In particular, we judiciously exploit the similarity between supersymmetric and Bateman's dual systems to provide a physical realization for the bosonic and fermionic fields in terms of bright and dark solitons.
\vskip 1.0 cm
\noindent{\bf\large 2. Evolution equations and their supersymmetric partners Introduction}
\vskip 0.3 cm
In this section we collect a number of well-known integrable nonlinear evolution equations which are related to the KdV equation and study their supersymmetric structure. We also present a similar treatment for the non integrable third-order RH equation.\\
\\
(i) Modified KdV equation: The nonlinear transformation of Miura or the so-called Miura transformation {[}11{]}
$$u=\xi_x+\xi^2\,\,,\,\,\,\xi=\xi(x,t)\eqno(10)$$
converts the KdV equation in $(8)$ into a modified KdV (mKdV) equation
$$\xi_t=-\xi_{3x}+6\xi^2\xi_x$$
For breviety, introducing  $\xi=u$ we write the mKdV equation in the form
$$u_t=-u_{3x}+6u^2u_x\,\,.\eqno(11)$$
This equation differs from the KdV equation only because of its cubic nonlinearity. It has many applicative relevance. For example, mKdV equation has been used to describe acoustic waves in anharmonic lattices and Alfv\'en waves in collisionless plasma.\\
\\
(ii)Hunter-Saxton equation: Consider the hereditary recursion operator $$R=c\partial^2 +\lambda(\partial u \partial^{-1} + u)\,\,,\eqno(12)$$
for the KdV hierarchy. Here $c$ and $\lambda$ are an arbitrary constants. This operator generates a hierarchy of integrable equations in which the first member is 
$$\frac{\partial u}{\partial t} = R u_x = cu_{xxx} + 3\lambda uu_x\,\,.\eqno(13)$$
 For $c=-1$ and $\lambda = 2$ we have the famous KdV equation. A second recursion operator can be extracted from $(12)$ by shifting  the function $u$ to $u+\gamma$, where $\gamma$ is a constant. We thus have
$$R(u+\gamma)= (c_1\partial^2 +\lambda(\partial u \partial^{-1} + u)) +(c_2\partial^{2} + 2\lambda\gamma)$$$$= R_1+R_2\,\,{\rm (say)}\,\,.\eqno(14)$$
In writing $(14)$ we have used $c=c_1+c_2$. It appears that the recursion operator $R = R_1R_2^{-1}$ generates new hierarchy of integrable equations
$$u_{t}=(R_1R_2^{-1})^n u_x\,\,.\eqno(15)$$
Assuming that $u=R_2\xi=c_2\xi_{xx}+2\lambda\gamma \xi$ the first member of the hierarchy is 
$$2\lambda \gamma \xi_t + c_2\xi_{xxt} =c_1\xi_{3x} + \lambda c_2\xi\xi_{3x} + 2\lambda c_2\xi_x\xi_{2x} +6 \lambda^2\gamma \xi\xi_x\,\,.\eqno(16)$$
For $\gamma = 0, c_1=0 , \lambda = -1$ and $\xi=u$ $(16)$ gives the Hunter-Saxton equation {[}7{]}
$$u_{xxt}+2u_xu_{2x}+uu_{3x}=0\,\,.\eqno(17)$$\\

(iii) Camassa-Holm equation: The choice $c_1=0,c_2=1,\gamma={1\over2}, \lambda=-1$  and $\xi=u$ $(16)$ leads to the well known Camassa-Holm equation {[}8{]}
$$u_t-u_{xxt}+3uu_x-2u_xu_{2x}-uu_{3x}=0\,\,.\eqno(18)$$
The CH equation describes the unidirectional propagation of shallow water waves over a flat bottom. It can also represent the geodesic flow on the Bott-Virasoro group.\\
\\
(iv) Rosenau-Hyman equation: The KdV and mKdV equations are quasi-linear. Here the dispersion produced is compensated by nonlinear effects resulting in the formation of exponentially localized solitons. As opposed to KdV and mKdV equations, the HS and CH equations are  fully nonlinear. These equations do not have exponentially localized soliton solutions. Instead they support peaked- and cusp-like solutions often called peakon and cuspon. Both HS and CH are integrable. Unlike HS and CH equations, the family of FNE equations proposed by Rosenau and Hyman {[}9{]} are nonintegrable but have solitory wave solutions with compact support. That is, they vanish identically outside the finite range. These solutions were given the name compactons. The compactons are robust within their range of existence. In contrast with the interaction of solitons supported by KdV and mKdV equations, the point at which two compactons collide is marked by the birth of a low-amplitude compacton-anticompacton pair. The third-order RH equation is given by 
$$u_t+3u^2u_x+6u_xu_{2x}+2uu_{3x}=0\,\,.\eqno(19)$$
Although $(11)$, $(17)$, $(18)$ and $(19)$ appear to be structurally different from the KdV equation, each of them can be supersymmetrized in a rather straightforward manner. In Table $1$ we present results for equations of the superfield and supersymmetric partners resulting from B-supersymmetrization of these equations. For completeness we also include results for the KdV equation.
\vskip 0.5 cm

\begin{center}
Table 1: Nonlinear evolution equations and their supersymmetric partners\\
\vskip 0.2 cm
{\small
{\begin{tabular}{|c|c|c|}
\hline
Evolution equations&Equations for Superfield &Supersymmetric partners\\

\hline
KdV: &&\\
$u_t+u_{3x}-6uu_{x}=0$&$F_t+D^6F-6DFD^2F=0$&$v_t+v_{3x}-6uv_{x}=0$ \\
mKdV:&&\\
$u_t+u_{3x}-6u^2u_{x}=0$&$F_t+D^6F-6(DF)^2D^2F=0$&$v_t+v_{3x}-6u^2v_x=0$ \\
Hunter-Saxton:&&\\
$u_{xxt}+uu_{3x}+2u_xu_{2x}=0$&$D^4F_t+DFD^6F+D^3FD^4F$&$v_{xxt}+uv_{3x}+u_xv_{2x}$\\
&$+D^2FD^5F=0$&$+u_{2x}v_{x}=0$\\
Camassa-Holm:&&\\
$u_t-u_{xxt}+3uu_x-uu_{3x}$&$F_t-D^4F_t+3DFD^2F-DFD^6F$&$v_t-v_{xxt}+3uv_x-uv_{3x}$\\
$-2u_xu_{2x}=0$&$-D^3FD^4F-D^2FD^5F=0$&$-u_xv_{2x}-u_{2x}v_{x}=0$\\
FNE:&&\\
$u_t+3u^2u_x+6u_xu_{2x}+$&$F_t+3(DF)^2D^2F+3D^3FD^4F$&$v_t+3u^2v_x+3u_xv_{2x} $   \\
$2uu_{3x}=0$&$+3D^2FD^5F+2DFD^6F=0$&$+3u_{2x}v_x+2uv_{3x}=0$\\
\hline
\end{tabular}}}
\end{center}

\vskip 0.5 cm
Looking closely into this table we see that the superfield equations of KdV, mKdV and FNE equations contain only $F_t$ as the time derivative parts of these equations. In contrast to this, the corresponding time derivative parts of HS and CH equations are $D^4F_t$ and $F_t-D^4F_t$. Understandably, the presence of $D^4F_t$ is associated with the mixed derivative term $u_{xxt}$ present in these equations. Note that, as expected, all fermionic fields satisfy linear equations.
\vskip 1.0 cm
\noindent{\bf\large 3. Fermionic equations from an action principle}
\vskip 0.3 cm
Our current understanding of all prototypical physical theories is largely based on the action principle as enunciated by Hamilton during mid 1930's. In the context of field theory, classical or quantum, the dynamical equations are obtained using the machineries of the Hamilton's variational principle. Thus it is expected that the bosonic and fermionic equations as given in Table 1 will follow from judicious use of the action principle. To achieve this we proceed by noting the following.\par A single evolution equation does not follow from an action principle {[}12{]}. When written in terms of the Casimir potential an evolution equation can either be Lagrangian or nonLagrangian. The example of a nonLagrangian evolution equation even in potential space is provided by Rosenau-Hyman equation {[}13{]}. Most of nonlinear evolution equation have at least one conserved density such that we can write these equations as
$$u_t+\frac{\partial\rho[u]}{\partial x}=0\,\,.\eqno(20)$$
Clearly, the KdV, mKdV and FNE equations are of the form $(20)$. The HS and CH equations can be recast in a similar form by introducing $n=u_{xx}$ and $m=u-u_{xx}$.\par We can make use of an elementary lemma to get a Lagrangian representation of $(20)$.
\vskip 0.5cm
\noindent{\bf Lemma 1.} There exists a prolongation of $(20)$ into another equation
$$v_t+{\delta\over\delta u}(\rho [u]v_x)=0\eqno(21)$$
with the variational derivative
$${\delta\over\delta u}=\sum_{k=0}^{n}(-1)^k{\partial^k\over\partial x^k}{\partial\over \partial u_{kx}},\,\,\,\,\,u_{kx}={\partial^ku\over\partial x^k} \eqno(22)$$
such that the system of equations follows from the action principle 
$$\delta \int {\cal L}\,\,dx dt=0\,\,.\eqno(23)$$
Here ${\cal L}$ stands for the Lagrangian density.
\vskip 0.5cm
\noindent{\bf Proof.}  For a direct proof of the lemma let us introduce ${\cal L}$ in the form
$$ {\cal L}={1\over 2}(vu_t-uv_t)-\rho[u]v_x\,\,.\eqno(24)$$ 
From $(23)$ and $(24)$ we obtain the Euler-Lagrange equations 
$${d\over dt}(\frac{\partial {\cal L}}{\partial v_t})-\frac{\delta {\cal L}}{\delta v}=0\eqno(25)$$
and 
$${d\over dt}(\frac{\partial {\cal L}}{\partial u_t})- \frac{\delta {\cal L}}{\delta u}=0\,\,.\eqno(26)$$
Using $(24)$ in $(25)$ and $(26)$ we obtain $(20)$ and $(21)$ respectively.
\vskip 0.5cm
\par Lemma 1 gives the  fermionic equations for the KdV and mKdV equation in a rather straightforward manner. Since the HS and CH equations involve mixed derivative like $u_{xxt}$, one needs to make use of a simple variant of Lemma 1 to construct the corresponding fermionic equations. To achieve this we write the CH equation as
$$m_t+\frac{\partial\rho[u]}{\partial x}=0\eqno(27)$$
with the associated equation 
$$\eta_t+{\delta\over\delta u}(\rho[u]v_x)=0\eqno(28)$$
where $\eta=v-v_{xx}$. The system of equations will follow from a Lagrangian density
$$ {\cal L}={1\over 2}(vm_t-mv_t)-\rho[u]v_x\eqno(29)$$
via the Euler-Lagrange equation
$${d\over dt}{d^2\over dx^2}(\frac{\partial {\cal L}}{\partial \phi_{xxt}})+{d\over dt}(\frac{\partial {\cal L}}{\partial \phi_t})-\frac{\delta {\cal L}}{\delta \phi}=0\eqno(30)$$ 
obtained from $(23)$ and $(29)$. Here $\phi$ is either $u$ or $v$. A similar treatment also applies for the HS equation. \par Application of Lemma 1 to the FNE equation gives the superpartner
$$v_t+3u^2v_x+2uv_{3x}=0\,\,.\eqno(31)$$
This equation does not agree with the result given in Table 1 presumably because $(19)$ is nonLagrangian even in the potential representation. In the recent past one of us (BT) {[}13{]} found that the Helmholtz solution of the inverse problem {[}14{]} can be judiciously exploited to construct Lagrangian systems from $(19)$, which support compacton solutions. For example, the equation 
$$u_t+3u^2u_x+8u_xu_{2x}+4uu_{3x}=0\eqno(32)$$
when written in terms of Casimir potential $w(x,t)=\int_x^\infty u(y,t)dy$ was found to result from an action principle and support compacton solution. The fermionic equation corresponding to $(32)$ can be obtained as
$$v_t+3u^2v_x+4u_xv_{2x}+4u_{2x}v_x+4uv_{3x}=0\,\,.\eqno(33)$$
It is easy to verify that $(33)$ can also be obtained by the use of Lemma 1. Thus we infer that all fermionic equations resulting from B-supersymmetrization of Lagrangian system of nonlinear evolution equations can be obtained by using our Lemma $1$.\par About a decade ago Kaup and Malomed {[}15{]} sought an application of the variational principle to nonlinear field equations involving dissipative terms. These authors demanded that the Lagrangian density
$${\cal L}=v(x,t)\,\,\times\,\,({\rm  equation \,for}\,u(x,t))\,\,.\eqno(34)$$
should be minimize the action functional. An important consequence of writing $(34)$ is that in the presence of the auxiliary field $v(x,t)$ the resulting set up Lagrangian symmetries coincides with the symmetries of the evolution equation for $u(x,t)$. In point mechanics this type of generalization for the traditional Noetherian symmetry was introduced by Hojman {[}16{]}.\par For the KdV equation $(34)$ reads
$${\cal L}=v\left(u_t+6uu_x+u_{3x}\right)\,\,.\eqno(35)$$
Clearly, for ${\cal L}$ in $(35)$ the Euler-Lagrange equation in $v(x,t)$ gives the KdV equation and more significantly we get the corresponding fermionic equation from the Euler-Lagrange equation for $u(x,t)$. Starting from $(34)$ we can verify the results given in Table 1 for the mKdV, HS and CH equations. Consistently with our previous analysis the use of $(34)$ does not reproduced the tabulated result for the RH equation. Rather we find that $(33)$ is the supersymmetric partner of $(32)$ which is Lagrangian in the potential space. Thus there must exist a relation between the Lagrangian densities in $(24)$ and $(34)$, For the KdV field the gauge terms $\frac{d}{dt}\left({1\over 2}uv\right)+\frac{d}{dx}\left(3u^2v\right) +\frac{d}{dx}\left(u_{2x}v\right)$  can be added to $(24)$ to get $(35)$. A similar conclusion also holds good for other Lagrangian equations in Table 1 implying that the Lagrangians in $(24)$ and $(34)$ are gauge equivalent.
\vskip 1.0 cm
\noindent{\bf\large 4. Conclusion}
\vskip 0.3 cm
Nonlinear evolution equations do not admit Lagrangian representation in terms of the field variables because the Fr\'echet derivative $D_P$ of any equation written as $u_t=P[u]$ is non self-adjoint such that the Helmholtz theorem {[}14{]} becomes inapplicable. Similar nonLagrangian systems also occur in point mechanics. For example, it is well known that there is no direct method of applying variational principles to nonconservative systems, which are characterized by friction or other dissipative processes. In fact, one can not write a time-independent Lagrangian even for a simple system like the damped Harmonic oscillator represented by
$$\ddot q_1+\lambda \dot q_1+\omega^2 q_1=0\eqno(36)$$
since it is non self-adjoint. Here $\lambda$ is the frictional coefficient and $\omega$, the natural frequency of the oscillator. Bateman {[}10{]} showed that while $(36)$ is non Lagrangian, the variational principle can be applied to a dual system consisting of $(36)$ and the equation
$$\ddot q_2-\lambda \dot q_2+\omega^2 q_2=0\,\,.\eqno(37)$$
Equations $(36)$ and $(37)$ can be obtained by using the Lagrangian
$$L=q_2(\ddot q_1+\lambda \dot q_1+\omega^2 q_1)\eqno(38)$$
in the Euler-Lagrange equations
$$\frac{\partial L}{\partial q_2}=0\eqno(39)$$
and
$$\frac{\partial L}{\partial q_1}-\frac{d}{dt}\left(\frac{\partial L}{\partial \dot q_1}\right)+\frac{d^2}{dt^2}\left(\frac{\partial L}{\partial \ddot q_1}\right)=0\,\,.\eqno(40)$$
Note that $L$ in $(38)$ can be converted to the well-known Morse-Feshbach {[}17{]} Lagrangian by adding appropriate gauge terms.\par Equation $(36)$ represents the motion of a simple pendulum embedded in a fluid which opposes the motion through frictional forces proportional to the velocity. As a result the energy is drained out of the system. The complementary equation $(37)$ represents a second physical system which absorbs energy dissipated in the first such that $(36)$ and $(37)$ taken together can be viewed as a conservative system to have an explicitly time independent Lagrangian representation. Thus the physical interpretation of Bateman dual system is fairly simple and straightforword. This is, however, not the case with the pair of equations obtained by B-supersymmetrization and finally shown to form a Lagrangian system. Nonintegrability of these equations {[}2,5{]} might be one the reasons for that.\par The single soliton solution of the KdV equation in $(8)$ is given by $u=-2k^2\, sech^2\,k(x-4k^2t)$. The solution has amplitude $-2k^2$ and moves to the right with speed $4k^2$. For this value of $u$ we have found $v$ in $(8^\prime)$ as $v=tanh^2\,k(x-4k^2t)$. The vave represented by $v$ is of constant amplitude and moves with same velocity as that of $u$. Moreover, $|u|$  with a peak while $|v|$ is a notch. Looking from this point of view $u$ and $v$ may be regarded as bright and dark soliton respectively. The bosonic and fermionic fields of $(32)$ and $(33)$ are associated with compaction and anticompacton expressed in terms of appropriate Jacobi elliptic function. This kind of realization for bosonic and fermionic fields marks a point of departure from our understanding of quantum field theory in that the particles and antiparticles are not superpartner.

\vskip 0.5 cm
{\bf Acknowledgements} \\
\\
This work is supported by the University Grants Commission, Government of India, through grant No. F.32-39/2006(SR).
\vskip 1.2 cm
{\bf References}
\vskip 0.8 cm

{[}1{]} Wess J and Zumino B 1974 {\it Nucl. Phys. {\bf B70}}39\\

{[}2{]} Mathieu P 1988 {\it J. Math. phys.} {\bf 29} 2499-2506 \\

{[}3{]} Korteweg D J and de Vries G 1895 {\it Phil. Mag.} {\bf 39} 422\\

{[}4{]} Zabusky N J and Kruskal M D  1965 {\it Phys. Rev. Lett.} {\bf 15} 240\\

{[}5{]} Das A arXiv:hep-th/0110125v1 15 Oct 2001\\

{[}6{]} Calogero F  and Degasperis A 1982 Spectral Transform and Soliton (New York: North-Holland Publising Company)\\

{[}7{]} Hunter J and Zheng Y 1994 {\it Physica D} {\bf 79} 361\\

{[}8{]} Camassa R and Holm D 1993 {\it Phys. Rev. Lett.} {\bf 70} 1661 \\

{[}9{]} Rosenau P and  Hyman J M 1993 {\it Phys. Rev. Lett.} {\bf 70} 564 \\

{[}10{]} Bateman H 1931 {\it Phys. Rev.} {\bf 38} 815\\

{[}11{]} Miura R M 1968 {\it J. Math. Phys.} {\bf 9} 1202 \\

{[}12{]} Olver P J 1993 Application of Lie Groups to Differential Equation (New York: Springer-Verlag) \\

{[}13{]} Talukdar B, Ghosh S, Samanna J and Sarkar P 2002 {\it European Physical Journal D} {\bf 21}, 105 ; Ghosh S, Samanna J and Talukdar B 2003 {\it Pramana-Journal of Physics} {\bf 61} 93 ; Ghosh S, Das U and Talukdar B 2005 {\it Int. J. Theor. Phys.} {\bf 44} 363 \\

{[}14{]} Helmholtz H 1887 {\it J. Reine Angew. Math.} {\bf 100} 137\\

{[}15{]} Kaup D J and Malomed B A 1995 {\it Physica D} {\bf 87} 155\\

{[}16{]} Hojman S 1984 {\it J. Phys. A} {\bf 17} 2399\\

{[}17{]} Santilli R M 1978 Foundations of Theoretical Mechanics I (New York: Springer-Verlag)
\end{document}